\documentclass{aa}
\usepackage{psfig}
\usepackage{graphicx}
\def\la{\;
\raise0.3ex\hbox{$<$\kern-0.75em\raise-1.1ex\hbox{$\sim$}}\; }
\def\ga{\;
\raise0.3ex\hbox{$>$\kern-0.75em\raise-1.1ex\hbox{$\sim$}}\; }
\newcommand{\zabs}{$z_{\rm abs}\,$}
\newcommand{\zem}{$z_{\rm em}\,$}
\newcommand{\kms}{km~s$^{-1}\,$}
\newcommand{\cm}{cm$^{-2}\,$}
\newcommand{\cmm}{cm$^{-3}\,$}
\newcommand{\nH}{$n_{\rm H}$}
\newcommand{\Tkin}{$T_{\rm kin}$}
\begin{document} 
\title{Detection of molecular hydrogen at \textit{z}\,=\,1.15 
toward HE~0515--4414\thanks{Based on 
observations with the NASA/ESA
Hubble Space Telescope, obtained at the Space Telescope Science Institute,
which is operated by Aura, Inc., under NASA contract NAS 5-2655; and on
observations collected at the VLT/Kueyen telescope ESO, Paranal,
Chile, programme ID 066.A-0212 }
}
\author{
D. Reimers\inst{1}
\and R. Baade\inst{1}
\and R. Quast\inst{1}
\and S. A. Levshakov\inst{2}
}
\offprints{S. A. Levshakov}
\institute{
Hamburger Sternwarte, Universit\"at Hamburg,
Gojenbergsweg 112, 21029 Hamburg, Germany
\and
Department of Theoretical Astrophysics,
Ioffe Physico-Technical Institute,
194021 St. Petersburg, Russia
}
\date{Received 00  / Accepted 00 }
\abstract{A new molecular hydrogen cloud is found in the sub-damped 
Ly$\alpha$ absorber
[$\log N$(\ion{H}{i}) = $19.88\pm0.05$]
at the redshift \zabs = 1.15 toward the bright quasar HE~0515--4414 (\zem = 1.71). 
More than 30 absorption features in the Lyman band system of H$_2$
are identified in the UV spectrum of this quasar
obtained with the Space Telescope Imaging Spectrograph (STIS)
aboard the {\it Hubble Space Telescope}.
The H$_2$-bearing cloud shows a total H$_2$ column density
$N$(H$_2$) $= (8.7^{+8.7}_{-4.0})\times10^{16}$ \cm\, and
a fractional molecular abundance 
$f_{{\rm H}_2} = (2.3^{+2.3}_{-1.1})\times10^{-3}$
derived from the H$_2$ lines arising from the $J = 0-5$ rotational levels
of the ground electronic vibrational state.
The estimated rate of photodissociation at the cloud edge 
$I_0 \la 1.8\times10^{-8}$ s$^{-1}$ is much higher
than the mean Galactic disk value, $I_{\rm MW} \sim 5.5\times10^{-11}$ s$^{-1}$. 
This may indicate an enhanced star-formation activity in the
$z = 1.15$ system as compared with molecular clouds 
at $z \sim 3$ where $I \sim I_{\rm MW}$.
We also find a tentative evidence that the formation rate 
coefficient of H$_2$ upon grain surfaces  at $z = 1.15$
is a factor of 10 larger  
than a canonical Milky Way value, $R_{\rm MW} \approx 3\times10^{-17}$ 
cm$^3$~s$^{-1}$.
The relative dust-to-gas ratio estimated from the [Cr/Zn] ratio
is equal to $\tilde{k} = 0.89\pm0.19$ (in units of the mean Galactic disk value),
which is in good agreement with a high molecular fraction in this system.
The estimated line-of-sight size of $L \sim 0.25$ pc may imply that the H$_2$
is confined within small and dense filaments embedded in a more rarefied gas giving
rise to the $z = 1.15$ sub-damped Ly$\alpha$ absorber.
\keywords{Cosmology: observations ---
Quasars: absorption lines ---
Quasars: individual: HE~0515--4414}
}
\authorrunning{D. Reimers et al.}
\titlerunning{H$_2$ at $z=1.15$ toward HE~0515--4414}
\maketitle

\section{Introduction}

The most abundant interstellar molecule in the universe, H$_2$, is currently observed not only
in the Milky Way disk (e.g., Rachford et al. 2002) and halo (e.g., Richter et al. 2003a), 
but also in the Magellanic Clouds (e.g., Tumlinson et al. 2002)
and in more distant regions
of the universe such as intervening damped Ly$\alpha$ absorbers (DLAs)
seen in spectra of background quasars (QSOs).
The DLAs are the systems with neutral hydrogen column densities 
$N$(\ion{H}{i}) $> 2\times10^{20}$ \cm. They are believed to originate
in protogalactic disks (Wolfe et al. 1995). The systems with lower hydrogen column densities,
$10^{19}$ \cm\, $\la N$(\ion{H}{i}) $\leq 2\times10^{20}$ \cm, are formally called sub-DLAs.
The sub-DLAs may also be related to intervening galaxies.
At the moment there are known 9  
molecular hydrogen systems detected in DLAs and sub-DLAs in the redshift range
from \zabs = 1.96 to 3.39 (see Table~1).

In this paper, we present results from the analysis of
a new 10th H$_2$ system detected at \zabs = 1.15 in the sub-DLA toward 
the bright quasar HE~0515--4414.
This is the first detection of H$_2$ with the STIS/HST at an intermediate redshift -- a
cosmological epoch when an enhanced star formation rate (SFR)
is observed in young galaxies.
The SFR shows a peak at $z \sim 1$ over the redshift interval
$0 \leq z \la 3$ (see, e.g., Hippelein et al. 2003 and references therein).

Molecular hydrogen, being an important coolant for gravitational collapse of gas clouds
at $T \sim 10^3$ K, is known to play a central role in star formation processes, and thus
one may expect that  the SFR and the fractional abundance of
H$_2$ are correlated.
Studying H$_2$-bearing cosmological clouds leads to better understanding of the physical
environments out of which first stellar populations were formed. 

\begin{table*}
\centering
\caption{Molecular hydrogen abundances 
and dust contents $\tilde{k}$ in cosmological H$_2$-bearing clouds}
\label{tab1}
\begin{tabular}{l c r @{$\pm$} l c c c l}
\hline
\noalign{\smallskip}
\multicolumn{1}{c}{QSO} & \zabs & 
\multicolumn{2}{c}{$\log N$(\ion{H}{i})} &   
$\log N$(H$_2$) &   
$\log f_{{\rm H}_2}$ & $\tilde{k}^\ast$ &   
H$_2$ detection \\
\noalign{\smallskip}
\hline
\noalign{\smallskip}
$0515-4414$ & 1.151 &19.88&0.05 & $16.94^{+0.23}_{-0.41}$ 
& $-2.64^{+0.30}_{-0.27}$ & $0.89\pm0.19$ & this paper\\
\noalign{\smallskip}
$0551-366$ & 1.962 & 20.50&0.08 & $17.42^{+0.63}_{-0.90}$
& $-2.78^{+0.64}_{-0.90}$ & $2.12\pm0.48$ & Ledoux et al. 2002\\ 
\noalign{\smallskip}
$0013-004^a$ & 1.973 & 20.70&0.05 & $19.84^{+0.10}_{-0.10}$ 
& $-0.66^{+0.10}_{-0.10}$ & $0.52\pm0.10$ & Ge \& Bechtold 1997\\
\noalign{\smallskip}
$1444+014$ & 2.087 &  20.07&0.07 & $18.30^{+0.37}_{-0.37}$
& $-1.48^{+0.38}_{-0.38}$  & $0.77\pm0.30$ & Ledoux et al. 2003\\
\noalign{\smallskip}
$1232+082$ & 2.338 & 20.90&0.10$^b$ &  $\geq 17.19^c$ &  $\geq -3.41^c$ & $0.35\pm0.12$ & Ge et al. 2001\\
\noalign{\smallskip}
$0841+129$ & 2.374 & 20.95&0.10$^c$ & $14.56^\dagger$ & $<-5.98$ & $0.024\pm0.005$ & Petitjean et al. 2000\\
\noalign{\smallskip}
$0405-443$ & 2.595 & 20.90&0.10 & $18.16^{+0.21}_{-0.06}$ & $-2.44^{+0.23}_{-0.12}$ & $0.17\pm0.06$
& Ledoux et al. 2003\\
\noalign{\smallskip}
$0528-250$ & 2.811 & 21.35&0.10$^{d,e}$ & $^c18.22^{+0.23}_{-0.16}$ & $^c-2.83^{+0.25}_{-0.19}$ &
$0.29\pm0.08$ & Levshakov \& Varshalovich 1985\\
\noalign{\smallskip}
$0347-3819$ & 3.025 & 20.63&0.01 & $14.61^{+0.02}_{-0.02}$ & $-5.71^{+0.02}_{-0.02}$ & $0.068\pm0.013$
& Levshakov et al. 2002 \\
\noalign{\smallskip}
$0000-2620$ & 3.390 & 21.41&0.08$^f$ & $^g13.94^\dagger$ & $\simeq -7.2$ & $\leq 0.002$ &
Levshakov et al. 2000\\
\noalign{\smallskip}
\hline
\noalign{\smallskip}
\multicolumn{8}{l}{Note: Column densities $N$(\ion{H}{i}) and $N$(H$_2$) are given in \cm;
$^\dagger$tentative H$_2$ identification; $^\ast$photospheric solar}\\  
\multicolumn{8}{l}{abundances for Cr and Zn are taken from Grevesse \& Sauval (1998), for
Mg and Fe from Holweger (2001), }\\
\multicolumn{8}{l}{see Eq.(\ref{eq:E3}).}\\
\multicolumn{8}{l}{$^a$High resolution UVES data reveal a few H$_2$ subcomponents
spread over $\sim 700$ \kms (Petitjean et al. 2002);}\\
\multicolumn{8}{l}{$^b$Srianand et al. 2000; $^c$Ledoux et al. 2003; $^d$Levshakov \& Foltz 1988:
$^e$M$\o$ller \& Warren 1993;}\\
\multicolumn{8}{l}{$^f$Prochaska \& Wolfe 1999; $^g$Levshakov et al. 2001. }
\end{tabular}
\end{table*}

\begin{figure}   
\vspace{0.0cm}
\hspace{0.5cm}\psfig{figure=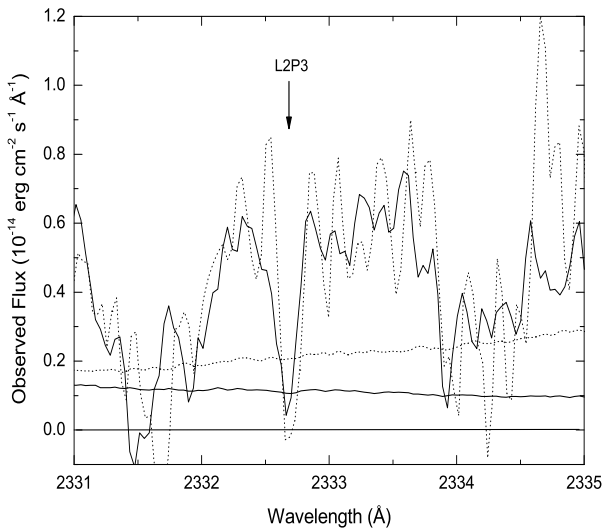,height=7.0cm,width=8.0cm}
\vspace{0.0cm}
\caption[]{
Typical example for a relative wavelength shift between
successive orders. The H$_2$ L2-0 P(3) line shows a difference of
$\Delta\lambda \sim0.04$ \AA\, ($\Delta v \sim5$ \kms). 
The lines at the bottom indicate the noise level.
}
\label{fig1}
\end{figure}

\begin{figure*}   
\vspace{0.0cm}
\hspace{-0.8cm}\psfig{figure=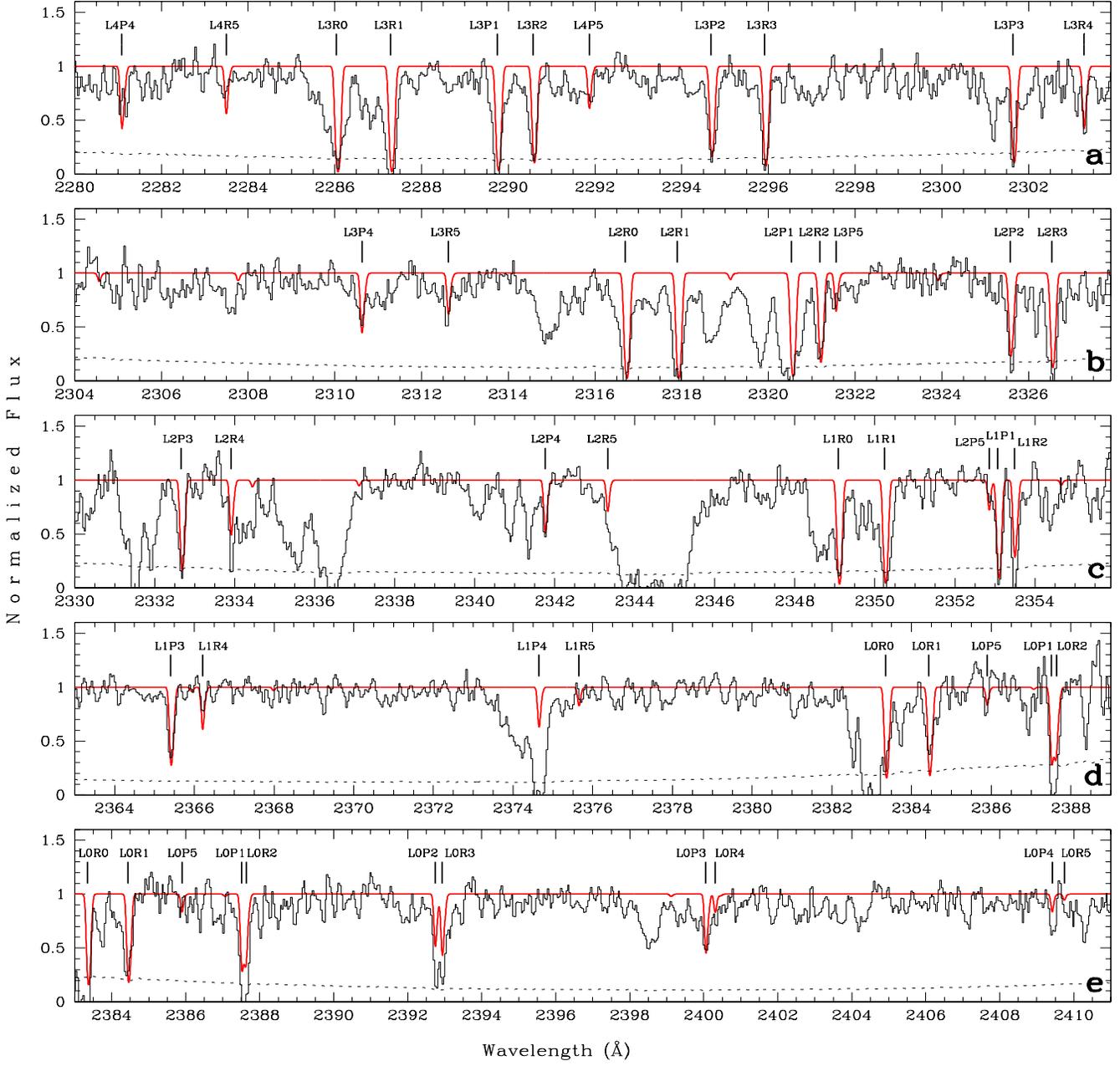,height=18.0cm,width=20.0cm}
\vspace{-1.0cm}
\caption[]{Continuum-normalized STIS spectra (histograms) of the quasar HE~0515--4414 
(individual echelle orders marked by {\bf a}, {\bf b}, {\bf c}, {\bf d}, and {\bf e}) 
and over-plotted synthetic H$_2$ profiles (smooth lines) arising from the rotational levels
$J = 0$ to $J = 5$ of the lowest vibrational level $v = 0$ in the ground electronic state
$X^1\Sigma^+_g$.
The noise level is shown by the dashed lines in each panel. 
The identified  H$_2$ lines are calculated for a two-component model with the components 
located at the measured redshifts of the \ion{C}{i} lines $z_1 = 1.15079$ and
$z_2 = 1.15085$, having the widths $b_1 = 2.0$ \kms, $b_2 = 3.5$ \kms\, 
(Quast et al. 2002), and a column density ratio $N_2/N_1 = 0.1$. 
}
\label{fig2}
\end{figure*}

\section{Observations}

Spectral data of the quasar HE~0515--4414 (\zem = 1.71, $V = 14.9$; Reimers et al. 1998) 
in the UV range were obtained 
with the HST/STIS (Reimers et al. 2001).
The medium resolution NUV echelle mode (E230M) and a $0.2''\times0.2''$
aperture provides a resolution power of 
$\lambda/\Delta\lambda \sim 30,000$ (FWHM $\simeq 10$ \kms). 
The overall exposure time was 31,500 s.
The spectrum covers the range between 2279~\AA\, and 3080~\AA\,
where the signal-to-noise ratio (S/N) per resolution
element varies from S/N $\sim 20$ to $\sim 5$.
The data reduction was performed by the HST pipeline completed
by an additional inter-order background correction and by coadding the
separate sub-exposures.

The spectral portion where the H$_2$ lines occur suffers from a poor S/N
ratio ($\la 20$).
An additional problem arises from the limited wavelength
accuracy. The MAMA detectors produce an absolute wavelength definition
between 0.5-1.0 pixel ($2\sigma$ limit as given by Brown et al.  2002). 
For our data 1 pixel corresponds to 0.038 \AA. The spectral
overlap of successive echelle orders allows to examine the wavelength
errors from order to order. Using well-defined line profiles we find
relative wavelength shifts of 0.02-0.05 \AA\, (see Fig.~1 for an
example).

Additional echelle spectra of HE 0515--4414 were obtained during
ten nights between October 7, 2000 and January 2, 2001 using
the UV-Visual Echelle Spectrograph (UVES) installed
at the VLT/Kueyen telescope. These observations were carried
out under good seeing conditions
(0.47--0.70 arcsec) and a slit width of 0.8 arcsec giving
the spectral resolution of $\lambda/\Delta\lambda \sim 55,000$
(FWHM $\simeq 6$ \kms).
The VLT/UVES data have a very high S/N ratio
($\simeq$ 50--100 per resolution element) which allows 
us to detect weak absorption features.

The high resolution VLT/UVES data reveal two narrow components
in the fine-structure \ion{C}{i} lines associated with the sub-DLA at
\zabs = 1.15 (de la Varga et al. 2000). The stronger
component at \zabs = 1.15079 is separated from the weaker one
by $\Delta v \simeq 8$ \kms, and shows 2.8 times higher column density 
(Quast et al. 2002, hereafter QBR). 
Exactly at the redshift of 
\ion{C}{i} lines we identified more than 30
absorption features in the Lyman band system of molecular hydrogen
H$_2$.

\begin{figure}
\centering
\scalebox{0.98}{\includegraphics*[0pt,0pt][284pt,165pt]{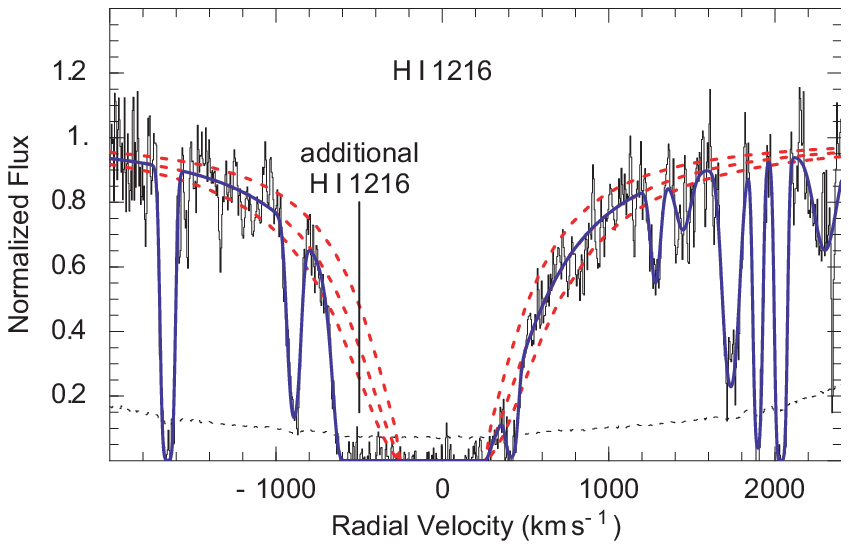}}
\caption[]{Part of the \ion{H}{i} Ly$\alpha$ forest showing the damped
Ly$\alpha$ system. The thin solid and dashed lines indicate the
continuum normalized flux and the noise level, respectively. The
thick solid curve represents the optimised Voigt profile model of the
spectrum. The thick dashed curves show the individual profiles of the
damped Ly~$\alpha$ line corresponding to 70, 100, and 130 percent of
the optimised column density $N$(\ion{H}{i}) $= 7.6\times10^{19}$ \cm.
The blue part of the damped line overlaps with some additional
\ion{H}{i} absorption components clearly identified by the presence of
many associated metal lines (Quast et al. 2003). The zero point of the
radial velocity corresponds to the redshift $z = 1.1508$.
}
\label{fig3}
\end{figure}

\begin{figure}
\vspace{-2.4cm}
\centering
\scalebox{0.98}{\includegraphics*[0pt,0pt][262pt,500pt]{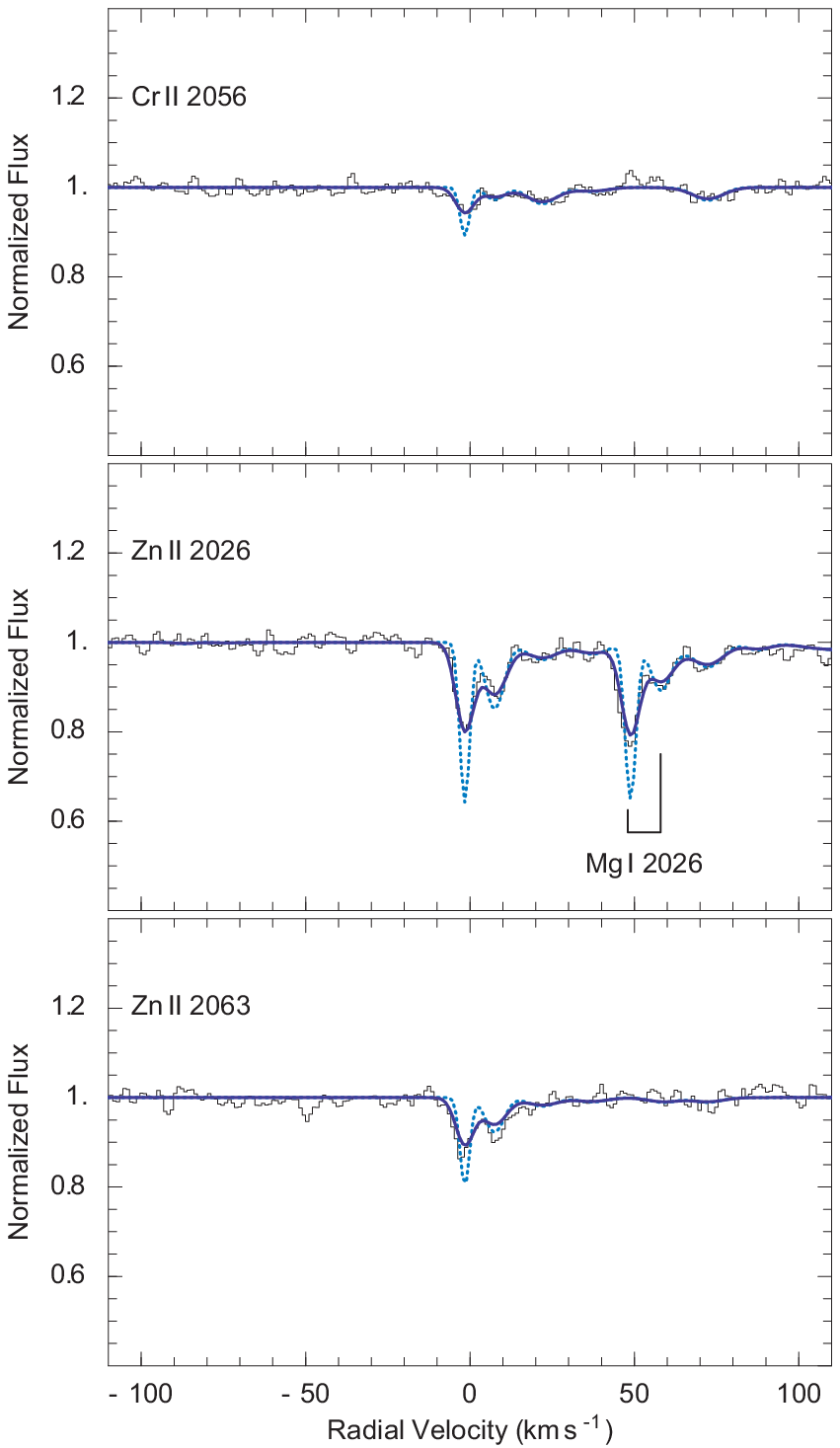}}
\caption[]{Parts of the UVES observations showing absorption arising
from the ions \ion{Cr}{ii} and \ion{Zn}{ii} (histograms). 
The solid and dashed lines represent our optimised model and its
deconvolution, respectively. \ion{Mg}{i} $\lambda 2026$ absorption
from the same sub-DLA is indicated.
The zero point of the radial
velocity corresponds to the redshift $z = 1.1508$.
}
\label{fig4}
\end{figure}

\section{Measurements}

In this section we describe the measurements
of the neutral hydrogen column density, metal and dust
content and the H$_2$ abundances in the \zabs = 1.15 sub-DLA.
These values are well known to be physically related.
The formation and maintenance of diffuse H$_2$ in the Milky
Way clouds is tightly correlated to the amount of interstellar
dust grains, which provide the most efficient H$_2$ formation
on their surfaces (see, e.g., Pirronello et al. 2000 and
references therein).

\subsection{Atomic hydrogen column density and metal abundance}

In order to estimate the column density of atomic hydrogen contained
in the sub-DLA, particular care has to be taken. Since the Doppler core of the
Ly$\alpha$ line is completely saturated, only the Lorentzian part
gives information about the line profile (Fig.~\ref{fig3}). Moreover, the
Lorentzian part is less pronounced than for typical DLAs and hence less
distinguishable from the quasar continuum. Therefore, we
simultaneously optimised the continuum 
and fitted the spectral features using
standard Voigt profile fitting technique. The
continuum is modelled as a linear function, and the
Voigt function is calculated using the pseudo-Voigt approximation
(Thompson et al. 1987).

Our optimized model (Fig.~\ref{fig3}) reveals some additional 
absorption in the blue part of the damped Ly$\alpha$ line at $\Delta v \simeq -420$ \kms.
This additional absorption is \ion{H}{i} Ly$\alpha$ which
is seen also in metal lines. The whole sub-DLA system is spread over 700 \kms\,
(Quast et al. 2003). 
This line together with 
other narrow absorption lines seen in the wings of the
damped Ly$\alpha$ were included in the Voigt fitting.

The derived column density of atomic hydrogen
in the sub-DLA is $\log N$(\ion{H}{i}) $= 19.88\pm0.05$,
where we estimated the standard deviation by varying the column density
until the resulting model profile is apparently inconsistent with the
observed data (Fig.~\ref{fig3}).

To measure the metal abundance in the main sub-component of the \zabs = 1.15 system
we used the \ion{Zn}{ii} $\lambda\lambda 2026,2063$ lines (Fig.~\ref{fig4}). 
The presence of dust grains in DLAs is usually estimated from the abundance ratio
[Cr/Zn]\footnote{$[X/Y] = (X/Y) - (X/Y)_\odot$, where $(X/Y)$ is the logarithmic value
of the element ratio by number without reference to the solar value. Photospheric
solar abundances $(X/Y)_\odot$ are taken from Grevesse \& Sauval (1998) and from
Holweger (2001).} 
assuming that Zn is undepleted (Pettini et al. 1994). 
In our high S/N spectrum, only a weak \ion{Cr}{ii} $\lambda 2056$ line was detected
at $\Delta v = 0$ \kms\, (Fig.~\ref{fig4}). Other \ion{Cr}{ii} lines ($\lambda\lambda
2062, 2066$) are too weak to be visible. Their oscillator strengths scale as
$f_{2056}:f_{2062}:f_{2066} = 1:0.74:0.50$ (Bergeson \& Lawler 1993).

\begin{table}
\centering
\caption{The H$_2$ column densities for different rotational levels
from the \zabs = 1.15 sub-DLA toward HE~0515--4414}
\label{tab2}
\begin{tabular}{c c c c}
\hline
\noalign{\smallskip}   
 Level & $\log N(J)$ & $\log N(J)$ & $\log N(J)$ \\
       & accepted & min & max \\
\noalign{\smallskip}
\hline
$J = 0$ & 16.47 & 16.00 & 16.70 \\
$J = 1$ & 16.60 & 16.00 & 16.85 \\
$J = 2$ & 15.85 & 15.70 & 15.95 \\
$J = 3$ & 16.00 & 15.90 & 16.18 \\
$J = 4$ & 15.00 & 14.85 & 15.30 \\
$J = 5$ & 14.48 & 14.30 & 14.60 \\
\noalign{\smallskip}
\hline
\end{tabular}
\end{table}

The column densities for \ion{Zn}{ii} and \ion{Cr}{ii} 
were calculated by Quast et al. (2003):
$\log N$(\ion{Cr}{ii}) = $12.01\pm0.09$ and
$\log N$(\ion{Zn}{ii}) = $11.99\pm0.02$. 
We used these values to estimate the dust-to-gas ratio in Sect.~4.2.

\subsection{{\rm H}$_2$ column densities}

Molecular hydrogen at \zabs = 1.15079 is detected in the $J = 0$ up to $J = 5$ rotational levels.
At a spectral resolution of $\sim10$ \kms\, it is not possible 
to resolve the internal structure in the H$_2$ lines 
observed in the \ion{C}{i} absorption (see Fig.~1 in QBR).
The H$_2$-bearing gas may also be distributed over a wider velocity range 
as compared with \ion{C}{i} which is easily ionised by UV photons in optically
thin zones.
However, for a good approximation one can assume that H$_2$
traces, in general, the volume distribution of \ion{C}{i} since such correlation is indeed
observed in the Milky Way (e.g., Federman et al. 1980).
Therefore, in our H$_2$ analysis we used a two-component model based on 
the observations of \ion{C}{i} by QBR. We note that the \ion{C}{i} data were obtained with
higher spectral resolution (FWHM $\simeq 5.5$ \kms) and considerably higher signal-to-noise
ratio (up to S/N $\simeq 130$ for the parts of the spectrum with \ion{C}{i} lines).

Panels {\bf a} -- {\bf e} in Fig.~2 present echelle orders of the STIS spectra 
of HE~0515--4414 (histograms) in the wavelength regions of the H$_2$ Lyman 0-0 to 4-0 bands, 
together with a two-component Voigt profile fit of the data (smooth lines). 
It is seen that some of the identified H$_2$ transitions are 
contaminated by the Ly$\alpha$ forest or blended with metals from different intervening
systems. 
This hampers significantly the measurements of 
accurate equivalent widths and their analysis through the curve of growth.
Moreover, the noise level (shown by the dashed line) is rather high for the available
STIS data and this may explain why some of H$_2$ features are inconsistent with others.
For instance, the observational profile of L4P4\footnote{Since all H$_2$ transitions
considered in the present paper arise from the ground electronic-vibrational state,
we use a short notation like L4P4 which means L4-0~P(4) in the standard form.}
in panel {\bf a} differs from those of L3R4 ({\bf a}), L3P4 ({\bf b}), L2R4 and L2P4 ({\bf c}), 
and L1R4 ({\bf d}). Relative strengths of the L0R0 and L0R1 lines from different
echelle orders (panels {\bf d} and {\bf e}) are not consistent (L0R0 is partly blended
with \ion{Fe}{ii} $\lambda 2383$ at $z = 0$). 
The apparent depths of the close pairs L0P2 + L0R3 ({\bf e}) and
L0P1 + L0R2 ({\bf e} and {\bf d}), 
as well as the single lines L1R2, L2R4 ({\bf c}) are deeper than those
calculated from the simultaneous fit to all H$_2$ lines.

Under these circumstances a standard $\chi$-square fitting cannot 
provide a statistically valuable measure of
goodness-of-fit. To estimate model parameters we required that the calculated
spectra were within 1 $\sigma$ uncertainty range for the majority 
of the unblended H$_2$ profiles 
or their unblended portions which match the data.

We tried to optimize a set of the H$_2$ column densities for the two-component model
with the components located at the measured redshifts 
of the \ion{C}{i} lines $z_1 = 1.15079$ and
$z_2 = 1.15085$. The broadening $b$-parameters 
were fixed at $b_1 = 2.0$ \kms, $b_2 = 3.5$ \kms\, 
(as deduced from the \ion{C}{i} lines by QBR), but a column density ratio between the
sub-components, $N_2/N_1$, was a free parameter ranging 
from 0.03 to 0.36 (the latter corresponds to the
\ion{C}{i} column density ratio found in QBR).

For a given H$_2$ component, 
the same $b$-parameter was used independently of the rotational level.
The column density in each $J$ level was derived from several calculations
of the Voigt profiles with a fixed 
value of $N_2/N_1$ and different $N(J)$ which match the observational
spectra.
The limiting values of $N(J)$ (an adjustable minimum and maximum)
were chosen to estimate the uncertainty interval for column densities.

All identified transitions from $J = 0$ and 1 are optically thick, but
the apparent central intensities of the L0R0 and L0R1 lines ({\bf d}) are not zero
(we consider the L0R0 line in panel {\bf e} as corrupted by a bad merging of different
spectra). These lines restrict $N(0)$ and $N(1)$ by, respectively, $5\times10^{16}$ and
$7\times10^{16}$ \cm\, at $N_2/N_1 = 0.03$.
On the other hand, we observe neutral carbon which is usually shielded
in molecular clouds from the background ionising UV radiation by the H$_2$  absorption
arising from the $J = 0$ and 1 levels. An essential shielding in H$_2$ lines
occurs when $N$(H$_2$) $\ga 10^{16}$ \cm. 
This gives us a hint at a possible range of $N(J=0,1)$. 

For the lower value of
$N_2/N_1 = 0.03$, the contribution from the second H$_2$ component is negligible, but the
synthetic profiles are systematically narrower as compared with the data. 
The presence of the second component is, therefore, important. On the other hand, the maximum
value of $N_2/N_1 = 0.36$ provides too wide synthetic profiles even with 
$N(0) = N(1) = 10^{16}$ \cm. We found that with $N_2/N_1 \sim 0.1$ an optimal set of the
H$_2$ column densities may be deduced. An example of such solution 
is shown in Fig.~2. The obtained results are given in Table~2.

\section{Discussion}

We investigate now the physical conditions in the \zabs = 1.15 H$_2$-bearing cloud
by considering the processes and parameters that balance the formation and dissociation
of molecular hydrogen. Since our observations show a relatively high metallicity in
this sub-DLA, [Zn/H] = $-0.49\pm0.10$ 
[i.e., $Z \sim \bigl ( \frac{1}{4}$\,-\,$\frac{2}{5} \bigr ) Z_\odot$],
and the dust content is approximately similar to the mean value for the cold gas in the
Galactic disk ($\tilde{k} \sim 0.9$), we consider catalytic reactions on the surfaces of
dust grains (Hollenbach \& Salpeter 1971) as the dominant H$_2$ formation process,
whereas ion-molecular gas phase reactions (Black 1978) are less efficient.

The measured column densities can be used to estimate the 
kinetic temperature, \Tkin, the gas density, \nH,
the photodissociation rate, $I$, and the rate of molecular formation on grains, $R$.
However, in view of the large uncertainties 
in the column densities $N$(0) and $N$(1),
we can only provide an order-of-magnitude estimate for these parameters. 

Another obstacle in the H$_2$ analysis is that the balance equation is related to the space
densities of \ion{H}{i} and H$_2$. In case of homogeneous clouds one can assume that
$n$(\ion{H}{i})/$n$(H$_2$) $\approx N$(\ion{H}{i})/$N$(H$_2$).
However, this assumption may not be correct 
for DLAs where multiphase structures and complex profiles
are usually observed. 
Observations show that with each step in increasing spectral resolution the profiles 
break up into subcomponents down to the new resolution limit.

The sub-DLA at \zabs = 1.15 reveals,
for example, transitions from neutrals 
and low ions
(as \ion{C}{i}, \ion{O}{i}, \ion{C}{ii}, \ion{Si}{ii}) 
to highly ionized ions 
(as \ion{C}{iv}, \ion{Si}{iv}) spread over $\Delta v \simeq 700$ \kms\,
(Quast et al. 2003) which implies that the neutral H$_2$-bearing 
cloud(s) is embedded in a lower density, higher temperature gas. 
Neutral hydrogen \ion{H}{i} 
can be spread over all gas phases that contain neutral gas with and without molecules.
The H$_2$ on the other hand may have a very inhomogeneous 
distribution in DLAs and concentrate in small clumps (Hirashita et al.
2003). Thus, only a fraction of the total \ion{H}{i} may be relevant to 
the formation of H$_2$. 
In our Galaxy, for instance, `tiny-scale atomic structures' (TSAS) 
and `small-area molecular structures ' (SAMS) 
in the ISM are observed (e.g., Lauroesch \& Meyer 1999;
Heithausen 2002). They show very high densities ($n_{\rm H} \sim 10^3-10^5$ \cmm) 
and very small sizes ($L \sim $ a few AU).

\begin{figure}   
\vspace{-1.0cm}
\hspace{-0.5cm}\psfig{figure=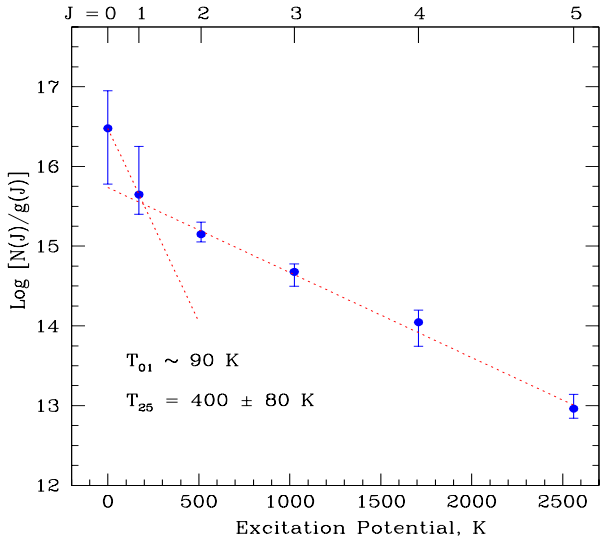,height=10.0cm,width=11.5cm}
\vspace{-2.0cm}
\caption[]{H$_2$ rotational excitation in the \zabs = 1.15 sub-DLA toward
HE~0515--4414. The logarithmic column densities $N(J)$, divided by the statistical
weight $g(J)$ for each state $J$, are plotted against the excitation potential.
The dashed lines represent fits from a theoretical Boltzmann distribution.
The rotational ground states, $J = 0$ and 1, fit on a line that is defined by
a temperature of $T_{01} = 90$ K. The excited levels, $J = 2-5$, fit on a line
with $T_{25} = 400\pm80$ K, possibly indicating excitation mechanisms 
through UV pumping and H$_2$ formation pumping.
}
\label{fig5}
\end{figure}

To take this uncertainty into account, a scaling factor $\phi \leq 1$ for the
\ion{H}{i} column density can be introduced. Following Richter et al. (2003b), who observed 
similar complex structures in the Milky Way halo molecular clouds, we define
\begin{equation}
\left[ \frac{n({\rm H\,\scriptstyle\rm I})}{n({\rm H}_2)} \right]_c =
\phi\,\frac{N({\rm H\,\scriptstyle\rm I})}{N({\rm H}_2)}\; ,
\label{eq:E1}
\end{equation}
where $c$ stands for the cloudlet(s) where H$_2$ is confined and $N$(\ion{H}{i}) is the total
neutral hydrogen column density along the sightline within the absorber.

\subsection{Kinetic temperature} 

The kinetic temperature of the gas is usually estimated through
the excitation temperature $T_{01}$ describing the 
relative populations of the $J=0$ and $J=1$ levels.
This temperature is proportional to the negative inverse of the slope of the excitation diagram
drawn through the points of the respective $J$ levels in a plot
$\log [N(J)/g(J)]$ versus $E(J)$ shown in Fig.~5. Here, $E(J)$ is the excitation energy
of the rotational level $J$ relative to $J=0$, and $g(J)$ is its statistical weight. 

Figure~5 shows that the value of $T_{01}$ 
is rather uncertain in our case because of large errors in
$N(0)$ and $N(1)$. Its mean value $T_{01} \simeq 90$~K corresponds 
to the excitation diagram shown by
the dotted line,  and its upper limit is about 270~K, which represents, 
probably, an upper limit for
\Tkin\, of the gas in the main sub-component of the \zabs = 1.15 system.

For levels with $J=2,3,4$, and 5 the accuracy of the column densities is higher and we find
$T_{25} = 400\pm80$~K. The difference between $(T_{01})_{\rm max}$ and $(T_{25})_{\rm min}$
is not significant and the points in Fig.~5 can be fitted, in principle, 
to a single excitation diagram.
But the previous analysis of the fine-structure level populations of \ion{C}{i},
where the most probable value for \Tkin\, of 240~K was found (QBR), indicates that these
two temperatures may not be equal.
This is also in line with results on the H$_2$ study in the Milky Way which revealed
that single excitation diagrams fit usually only
optically thin lines with $N$(H$_2$) $< 10^{15}$ \cm\, (Spitzer \& Cochran 1973).
For higher column densities, there is `bifurcation to two temperatures,
depending on the $J$ levels' (Jenkins \& Peimbert 1997).

\subsection{Fractional molecularization and dust content} 

According to our calculations presented in Sect.~3.1,
the total \ion{H}{i} column density in the main sub-component is
$N$(\ion{H}{i}) $= (7.6\pm0.9)\times10^{19}$ \cm. 
With $N$(H$_2$) $= (8.7^{+8.7}_{-4.0})\times10^{16}$ \cm, 
the ratio of H nuclei in molecules to the
total H nuclei is
\begin{equation}
f_{{\rm H}_2} = \frac{2N({\rm H}_2)}{N({\rm H})} = (2.3^{+2.3}_{-1.1})\times10^{-3}\; ,
\label{eq:E2}
\end{equation}
where $N$(H) = $N$(\ion{H}{i}) + 2$N({\rm H}_2)$.

Listed in Table~1 are the molecular hydrogen fractions in all known H$_2$ systems. 
The $f_{{\rm H}_2}$
values were derived in the standard way assuming the scaling factor $\phi = 1$. 
This may imply that
the listed molecular hydrogen fractional abundances are systematically underestimated. 

In Fig.~6 we compare these abundances with the dust-to-gas ratios, $\tilde{k}$, estimated from
(Vladilo 1998):
\begin{equation}
\tilde{k} = \frac{10^{[{\rm X}/{\rm H}]_{\rm obs}}}{f_{\rm X,ISM} - f_{\rm Y,ISM}}
\left( 10^{[{\rm X}/{\rm Y}]_{\rm obs}} - 1 \right)\; ,
\label{eq:E3}
\end{equation}
where X and Y are two heavy elements with different depletions in dust. For all systems except
Q~0405--443 and Q~1232+082 we used X = \ion{Zn}{ii} and Y = \ion{Cr}{ii} with their
fractions in dust $f_{\rm Zn,ISM} = 0.587\pm0.048$ and $f_{\rm Cr,ISM} = 0.920\pm0.010$
referring to the Galactic interstellar medium (Vladilo 2002, hereafter V02). 
Since column densities for \ion{Zn}{ii} and \ion{Cr}{ii} are not known for Q~1232+082,
we used in this case X = \ion{Mg}{ii},  Y = \ion{Fe}{ii}  and
$f_{\rm Mg,ISM} = 0.715\pm0.056$,  $f_{\rm Fe,ISM} = 0.939\pm0.004$\, from V02,
although Mg and Fe may have not the same nucleosynthetic history.
For Q~0405--443, the \ion{Cr}{ii} abundance is not available and we used \ion{Fe}{ii} instead
of \ion{Cr}{ii}. The errors of the $\tilde{k}$ values were calculated by applying error
propagation method to the column density measurements quoted in the literature.

Figure~6 demonstrates an 
apparent correlation between $f_{{\rm H}_2}$ and $\tilde{k}$ 
in the range $0 \leq \tilde{k} \leq 1$
which supports
the assumption that molecular hydrogen 
abundances in quasar absorbers are governed by the dust content
similar to that observed in the Galaxy. 
This conclusion rises the question: Why is the H$_2$ detection
in QSO absorbers in this case so rare (lower than 30\% according to Ledoux et al. 2003) ? 
Following Hirashita et al. (2003), we 
suppose that
a relative paucity of H$_2$ observations in DLAs may be caused by a bias against finding H$_2$
in dense molecular clumps that have 
a small angular extent and thus a small volume filling factor.
DLAs are mainly associated with diffuse clouds that have large volume filling factor and low
molecular fractions, 
but they may also contain a small size dense filaments like the above mentioned 
TSAS or SAMS.
Besides, 
low metallicity of the QSO absorbers can also significantly
suppress H$_2$ formation (Liszt 2002).

\begin{figure}   
\vspace{-6.0cm}
\hspace{-3.0cm}\psfig{figure=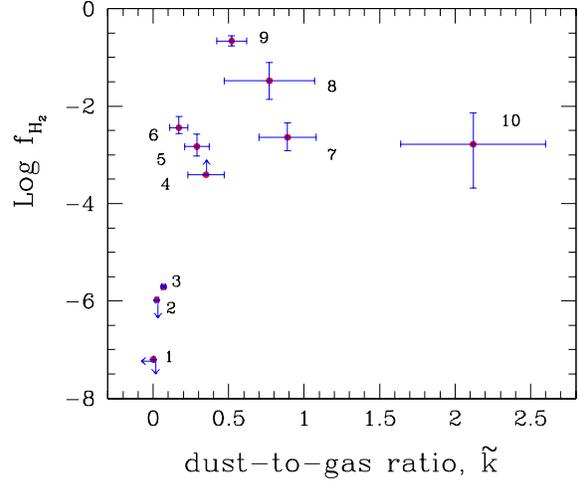,height=17.0cm,width=17.0cm}
\vspace{-4.5cm}
\caption[]{Relation between H$_2$ fractional abundance $f_{{\rm H}_2}$
plotted on a logarithmic scale and relative dust-to-gas ratio $\tilde{k}$
(with respect to the mean Milky Way value) in DLAs and sub-DLAs.
Indicated by numbers are the systems found in spectra of QSOs 0000--2620 (1),
0841+129 (2), 0347--3819 (3), 1232+082 (4), 0528--250 (5), 0405--443 (6),
0515--4414 (7), 1444+014 (8), 0013--004 (9), and 0551--366 (10).
A correlation between the two quantities 
in the range $0 \leq \tilde{k} \leq 1$
is apparent (a large shift of  
$\tilde{k}$ for Q~0551--366 is probably caused by  systematic errors in the
data obtained by Ledoux et al. 2002, see text).

}
\label{fig6}
\end{figure}

One point in Fig.~6 (Q~0551--366) shows an unrealistic high
dust-to-gas ratio, about 2 times the
Galactic value. This large value may be, probably, explained by
systematic errors in the measurement of the \ion{Zn}{ii} column density. 
For instance, the relative
abundance of Si, [Si/H] $=-0.42\pm0.11$, differs significantly
from [Zn/H] $=-0.08\pm0.12$ according to Ledoux et al. (2002).
The fraction of these elements in dust in the Milky Way is approximately identical,
$f_{\rm Si,ISM} =0.691\pm0.069$ 
and $f_{\rm Zn,ISM} =0.587\pm0.048$
(V02). 
At [Fe/H] $=-0.90\pm0.11$, Ledoux et al. measured
[Si/Fe] $\sim 0.5$ which is in line with other observations (see Fig.~1 in V02).
This means that [Zn/H] is 
most likely
overestimated in the \zabs = 1.962 system.

\subsection{{\rm H}$_2$ formation and photodissociation rates}

In equilibrium between formation on grains with rate coefficient $R$ (cm$^3$ s$^{-1}$)
and photodissociation with rate $I$ (s$^{-1}$), 
we may write that (Jura 1975b)
\begin{equation}
I n_{{\rm H}_2} = R n n_{\rm H} \approx 0.11\beta_0 n_{{\rm H}_2}\: ,
\label{eq:E4}
\end{equation}
where $n = n_{\rm H} + 2 n_{{\rm H}_2}$, and
$\beta_0$ is the photoabsorption rate depending on the
local UV radiation field (one may neglect the $J$ dependence of $\beta$ for
an optically thick cloud since the photoabsorption rates from the levels $J \geq 2$
are low).

To estimate 
the formation rate of H$_2$ upon grain surfaces $Rn_{\rm H}$, 
we use approximation described by Jura
(1975b). It assumes that $(i)$ the levels $J = 4$ and $J = 5$ are populated by direct
formation pumping and by UV pumping from $J = 0$ and $J = 1$, $(ii)$ the self-shielding
in the levels $J = 0$ and $J = 1$ is about the same, $(iii)$ the upper levels
$J = 4$ and $J = 5$ are depopulated by spontaneous emission (which is valid if 
$n_{\rm H} < 10^4$ \cmm). We do not consider additional rotational excitation of H$_2$
caused by a shock because restrictions on the gas density
($n_{\rm H} \sim 100$ \cmm) and kinetic temperature ($T_{\rm kin} \leq 240$ K)
set by the observations of \ion{C}{i}, \ion{C}{i}$^\ast$, and \ion{C}{i}$^{\ast\ast}$
(QBR) show that collisional excitation of the levels $J = 4$ and $J = 5$ is not
significant.
Using the cascade redistribution probabilities $p_{4,0} = 0.26$ and $p_{5,1} = 0.12$,
calculated by Jura (1975a), and assuming $T_{\rm kin} = 90$ K, we can re-write Eqs. (3a)
and (3b) from Jura (1975b) in the form
\begin{equation}
[Rn_{\rm H}]_c = 2.35\times10^{-9}\frac{N(4)}{N({\rm H\,\scriptstyle\rm I})}\phi^{-1},
\label{eq:E5}
\end{equation}
and
\begin{equation}
[Rn_{\rm H}]_c = 9.15\times10^{-9}\frac{N(5)}{N({\rm H\,\scriptstyle\rm I})}\phi^{-1}.
\label{eq:E6}
\end{equation}
By substituting numerical values in (\ref{eq:E5}) and (\ref{eq:E6}) we obtain,
respectively,
$[Rn_{\rm H}]_c =
(3.1^{+3.1}_{-1.0})\times10^{-14}\phi^{-1}$ s$^{-1}$ and
$(3.6\pm1.2)\times10^{-14}\phi^{-1}$ s$^{-1}$,
which are consistent in the range 
$[Rn_{\rm H}]_c = (4.1\pm1.7)\times10^{-14}\phi^{-1}$ s$^{-1}$,
and independent on the local value of $\beta_0$, as pointed out by Jura (1975b).

To estimate the photodissociation rate at the cloud surface $I_0$, the shielding
effect is to be taken into account. The shielding factor, $S$, depends on line overlap,
self-shielding of H$_2$, and continuum absorption. Lee et al. (1996) showed that these
various factors can be well represented by the H$_2$ column density. They calculated the
values of $S$ as a function of $N$(H$_2$) for a turbulent velocity of 3 \kms,
which suits well for our case.
From their Table~10 we find $S = (1.928^{+1.075}_{-0.656})\times10^{-3}$ for
$N$(H$_2$) = $(8.7^{-4.0}_{+8.7})\times10^{16}$ \cm, respectively.
This gives us a 
rough estimate of $I_0 = Rn_{\rm H}N$(\ion{H}{i})$/SN$(H$_2$) $= 1.8\times10^{-8}$
s$^{-1}$ (with the uncertainty of about 120\%), or
$\beta_0 \simeq 1.6\times10^{-7}$ s$^{-1}$.
The result obtained should be considered, however, as an upper limit on $\beta_0$
since in our estimations we assumed that the H$_2$ is one singe gas cloud.
If, in reality, the H$_2$ is inhomogeneously distributed among several cloudlets,
the value of $\beta_0$ should be lower. 

In the Milky Way, the mean value for
$\beta_{0,\rm halo} \approx 0.5\beta_{0,\rm disk} = 2.5\times10^{-10}$ s$^{-1}$ 
(e.g., Richter et al. 2003b)
and, thus, we may conclude that the H$_2$
in the \zabs = 1.15 sub-DLA is 
probably exposed to a radiation field with the intensity much higher than
the mean Galactic value.
For comparison, molecular clouds in the LMC and SMC also reveal 
10-100 times more intense UV radiation field than the Galactic one (Browning et al. 2003).
High rotational excitation of the H$_2$ observed in the LMC/SMC gas and
at \zabs = 1.15 is compared with Galactic data in Fig.~7. 
We note that the upper limit on the local UV field at \zabs = 1.15 of 
about 100 times the Galactic value was independently determined
from the analysis of the \ion{C}{i} fine-structure lines by QBR.

We can also estimate the photodissociation rate
produced by the intergalactic UV background (UVB) field
on the surface of a cloud (Hirashita et al. 2003):
\begin{equation}
I_{\rm UVB} \sim 1.38\times10^{-12} J_{21}\;,
\label{eq:E9}
\end{equation}
where $J_{21}$ (in units $10^{-21}$ erg \cm s$^{-1}$ Hz$^{-1}$ str$^{-1}$)
is the UVB intensity at 1 Ryd averaged for all the solid angle.
At \zabs = 1.15, $J_{21} \simeq 0.4$ (Haardt \& Madau 1996),
and thus $I_{\rm UVB}$ is of the order of
$5\times10^{-13}$ s$^{-1}$, which is about 4 orders of magnitude lower
as compared with the derived value of $I_0 \la 10^{-8}$ s$^{-1}$.
The presence of bright young stars in the \zabs = 1.15 sub-DLA 
is therfore required to maintain
this high value of $I_0$.

Star-formation activity is not, however, intense 
in the H$_2$-bearing clouds at higher redshift. For example,
the UV radiation fields in the \zabs = 3.025 H$_2$ absorber toward Q~0347--3819
and in the Galactic ISM are very much alike (Levshakov et al. 2002). Other examples
may be found in Ledoux et al. (2003).

\begin{figure}   
\vspace{0.0cm}
\hspace{-0.5cm}\psfig{figure=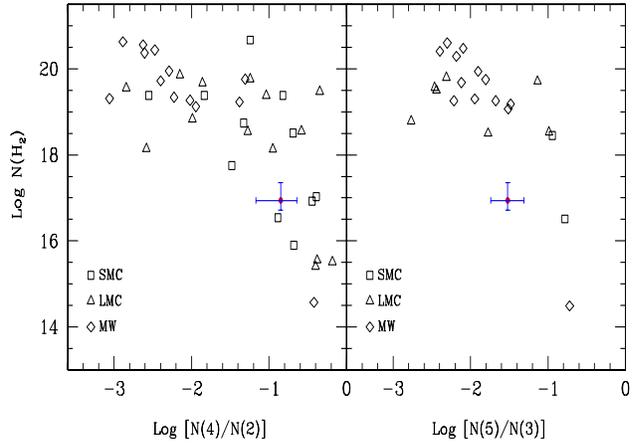,height=10.0cm,width=10.0cm}
\vspace{-4.0cm}
\caption[]{Total column density of H$_2$ vs. excitation ratios
$N(4)/N(2)$ and $N(5)/N(3)$, for LMC, SMC, Milky Way (data 
are taken from Browning et al. 2003), and sub-DLA at \zabs = 1.15
(dots with error bars). A systematically higher rotational excitation
is observed in the Magellanic Clouds and in the sub-DLA as compared
with Galactic data points. 
}
\label{fig7}
\end{figure}

\subsection{Gas density and {\rm H}$_2$ formation rate coefficient}

We now consider constraints on the volumetric gas density $n_{\rm H}$ and the H$_2$
formation rate coefficient $R$ stemming from the foregoing estimation of 
$[Rn_{\rm H}]_c \sim 4\times10^{-14}\phi^{-1}$ s$^{-1}$. 

A useful reference point is provided by the $J = 2$ level.
The population of this level is more directly affected by collisional processes, since
it  has a longer radiative lifetime as compared with $J=3$ and other levels\footnote{
$t_{J=2} = 3.4\times10^{10}$ s, and $t_{J=3} = 2.1\times10^{9}$ s (Turner et al. 1977)}.
The critical density, $n^{\rm cr}_{\rm H}$, at which the probability of collisional and
radiative de-excitation of $J=2$ are equal is 200 \cmm, if \Tkin\, = 100~K, and
$n^{\rm cr}_{\rm H} = 80$ \cmm, if \Tkin\, = 240~K (the collisional de-excitation
rate coefficients $q_{jj'}$ are taken from Forrey et al. 1997).
If $n_{\rm H} > n^{\rm cr}_{\rm H}$, collisional de-excitation becomes important. 

To estimate $n_{\rm H}$, the grain formation rate of H$_2$ should be known or vice versa.
$R$ is a complex function of the gas and dust temperature and other poorly known parameters
(Hollenbach \& McKee 1979). 
According to their calculations,
$R_{\rm MW} \approx 3\times10^{-17}$ cm$^3$~s$^{-1}$.
Using this value, we find $n_{\rm H} \ga 1300$ \cmm,
if $\phi \leq 1$. However,
for such large density, the $J=0$ and $J=2$ levels as well as the $J=1$ and $J=3$
levels should be in thermal equilibrium, which we do not observe. Moreover, the upper limit
on the gas density set from the excitation of \ion{C}{i} is 110 \cmm\, (QBR).
If we tentatively adopt $n_{\rm H} \sim 100$ \cmm, then $R \sim 10 R_{\rm MW}$. 

This value of $R$ is larger than predicted in theoretical calculations, 
but it is conceivable
that $R$ may vary in space since 
we observe considerable variations in the UV extinction among
different clouds. A  low grain formation rate of H$_2$ ($R \sim 0.1 R_{\rm MW}$) 
was, for instance, recently estimated in the LMC and SMC by Browning et al. (2003).
Although our high value of $R$ is consistent with the best determinations of upper
limits to $R$ toward $\gamma$ Peg, $\nu$ Sco, $\lambda $ Sco etc. (Jura 1974), 
we cannot without further observations 
conclude that this result is certain.
To test whether or not the grain formation rate coefficient $R$ 
exceeds the value of
$R_{\rm MW}$, a higher accuracy for the $N$(0) and $N$(1) column densities is needed
to verify that $T_{01} \neq T_{25}$.

For $n_{\rm H} \sim 100$ \cmm, the linear thickness of the H$_2$-bearing cloud is small,
$L \sim 0.25$ pc. Similar characteristics of molecular hydrogen small structures are found
in intermediate-velocity clouds (IVC) in the Milky Way halo (Richter et al. 2003b).

\section{Conclusions}

We have analyzed the HST/STIS and VLT/UVES spectra of the quasar HE~0515--4414
and deduced the physical properties of the H$_2$-bearing cloud embedded in the
sub-DLA at \zabs = 1.15.
The main conclusions are as follows:
\begin{enumerate}
\item In the STIS spectrum of HE~0515--4414
we have identified over 30 absorption features with the Lyman lines arising from the
$J = 0-5$ rotational levels of the ground electronic vibrational state of H$_2$.
These lines have exactly the same redshift as the fine-structure transitions
in \ion{C}{i} identified earlier by QBR in the UVES spectrum of this quasar.
\item We find a total H$_2$ column density of
$N$(H$_2$) $= (8.7^{+8.7}_{-4.0})\times10^{16}$ \cm\, and
a molecular hydrogen fraction of
$f_{{\rm H}_2} = (2.3^{+2.3}_{-1.1})\times10^{-3}$.
\item From the measured ratios [Cr/H] $= -1.54\pm0.11$  and  [Zn/H] $= -0.49\pm0.10$
we calculated the relative dust-to-gas ratio
$\tilde{k} = 0.89\pm0.19$ (in units of the mean Galactic disk value)
in the molecular cloud (the ratio [Cr/Zn] is usually used to indicate
the presence of dust in DLAs). The derived H$_2$ fractional abundance correlates with the
dust content showing increasing $f_{{\rm H}_2}$ with increasing
$\tilde{k}$ in cosmological molecular clouds. This indicates that catalytic reactions
on the surfaces of dust grains is the dominant H$_2$ formation process not only in the
diffuse clouds in the Milky Way but in the DLA systems as well.
\item Two excitation temperatures are required to describe the rotational
excitation of the H$_2$ gas: for the $J=0$ and $J=1$ levels we find $T_{01} \sim 90$~K
and the upper limit for the kinetic temperature of the gas of about 270~K.
For $J=2-5$ we derive $T_{25} = 400\pm80$~K.
\item From the relative populations of H$_2$ in the $J=4$ and 5 rotational levels we
estimated the rate of photodissociation at the cloud surface
$I_0 \la 1.8\times10^{-8}$ s$^{-1}$ which is much higher than the
mean Galactic disk value. Star-formation activity may be very intense in
the close vicinity of the H$_2$-bearing cloud, which is in line with the observed
high SFR in galaxies at intermediate redshift, $z \sim 1$. 
At higher redshift, $z \sim 3$, we do not observe such intense UV fields in the
molecular clouds.
\item We also find that the formation rate coefficient
of H$_2$ upon grain surfaces is 
probably
10 times higher as compared with the conventional value adopted for the Milky Way,
$R_{\rm MW } \approx 3\times10^{-17}$ cm$^3$ s$^{-1}$.
\item We find that in order to be consistent with the measurements of the
population ratios of the fine-structure levels in \ion{C}{i}, the 
gas density in the molecular cloud should be $\sim 100$ \cmm, which implies that
the line-of-sight size of this cloud is small, $L \sim 0.25$ pc.
Small sizes and a low level of the H$_2$ detections in the DLAs favour calculations of
Hirashita et al. (2003) who showed that the H$_2$ may have very inhomogeneous distribution
within these systems. The diffuse molecular hydrogen gas forms, probably, in small,
dense filaments during the cooling and fragmentation phase in DLAs.

\end{enumerate}

\begin{acknowledgements}
We thank our referee P. Richter for his comments and remarks.
S.A.L. acknowledges the hospitality of
Hamburger Sternwarte, Universit\"at Hamburg.
The work of S.A.L. is supported in part by the
RFBR grant No.~03-02-17522.
R.Q. is supported by the Verbundforschung of the BMBF/DLR under Grant No. 50 OR 9911~1.
\end{acknowledgements}

\end{document}